\documentclass{sf2a-conf2025}
\usepackage{graphicx}
\usepackage{hyperref}
\usepackage[]{natbib}  
\usepackage{epstopdf}

\def\BibTeX{{\rm B\kern-.05em{\sc i\kern-.025em b}\kern-.08em
    T\kern-.1667em\lower.7ex\hbox{E}\kern-.125emX}}
\bibpunct{(}{)}{;}{a}{}{,}  %%%%%%%%%%%%%  A&A bibliography style
\def\velvec{{\rm \mathbf v}}

\def\music{{\textsc{music}}~}
\begin{document}

\TitreGlobal{SF2A 2025}

%%-----------------------------------------------------------------
%%      the top matter
%%

\title{Unveiling stellar (and planetary) internal dynamics with the fully compressible \textsc{MUSIC} code}

\runningtitle{Fully compressible stellar hydrodynamics with MUSIC}

\author{A. Le Saux}\address{Université Paris-Saclay, Université Paris Cité, CEA, CNRS, AIM, Gif-sur-Yvette, F-91191, France.}

\author{I. Baraffe$^{2,}$}\address{University of Exeter, Physics and Astronomy, EX4 4QL Exeter, UK} \address{\'Ecole Normale Sup\'erieure, Lyon, CRAL (UMR CNRS 5574), Universit\'e de Lyon, France} 

\author{T. Guillet$^2$}

\author{J. Pratt$^{4,}$}\address{Lawrence Livermore National Laboratory, 7000 East Ave, Livermore, CA 94550, USA}\address{Department of Physics and Astronomy, Georgia State University, Atlanta, GA 30303, USA}

\author{T. Goffrey}\address{Centre for Fusion, Space and Astrophysics, Department of Physics, University of Warwick, Coventry CV4 7AL, UK}

\author{D. Vlaykov}\address{Department of Mathematics and Statistics, University of Exeter,
EX4 4QE, Exeter, UK}

\author{A. Morison}\address{Research Software Engineering, MVLS SRF, University of Glasgow,
Glasgow, UK}

\author{J. Morton$^2$}

\author{M. Stuck$^4$}%\address{Lawrence Livermore National Laboratory, 7000 East Ave, Livermore, CA 94550, USA}

\author{M. G. Dethero$^5$}%\address{Department of Physics and Astronomy, Georgia State University, Atlanta, GA 30303, USA}

\author{N. de Vries$^2$}

%% IF Author3 has the same affiliation than Author1:
%\author{C.\,E. Author3$^1$}

%% IF Author3 has its own affiliation:
%\author{C.\,E. Author3}\address{\'Ecole Normale Sup\'erieure, Lyon, CRAL (UMR CNRS 5574), Universit\'e de Lyon, France} 

%% IF Author3 has two affiliations, the one of Author1 and a second one:
%\author{the MUSIC Consortium}

%% Keep this line, even if the page will be settled afterwards.
\setcounter{page}{237}

%%-----------------------------------------------------------------

\maketitle
%%-----------------------------------------------------------------
%%        The abstract
%% 
%%  Warning!  within the abstract:
%%  - do not use macros. 
%%  - do not use commands like: \cite, \citet, \citep ... etc.

\begin{abstract}
Multidimensional hydrodynamical simulations have transformed the study of stellar interiors over the past few decades. Most codes developed during that time use the anelastic approximation, which fixes the thermal structure of simulations and filters out sound waves. Many of them also use explicit time integration, which imposes severe constraints on the time step of the simulations.
In this context, \music is developed to overcome these limitations. Its main scientific objective is to improve the phenomenological approaches used in 1D stellar evolution codes to describe major hydrodynamical and MHD processes. Here, we review recent applications of the \music code, that focus mainly on convection, convective boundary mixing and waves in stars that possess convective cores, shells and/or envelopes.
\end{abstract}

%% Insert the keywords (to appear in the ADS indexing)
%% Keywords must be separated by a comma
\begin{keywords}
Stars: interiors, Waves, Convection, Fluid dynamics, Software: simulations; Stars: oscillations
\end{keywords}

%%-----------------------------------------------------------------

\section{Introduction}
%%---------------------

Understanding the internal dynamics of stars has been a major challenge in astrophysics for a long time. Stellar evolution models are highly dependent on the transport of energy, angular momentum, and chemical elements through processes such as convection, overshooting, rotational mixing, and wave-induced transport. Traditionally, these processes have been modelled in one-dimensional stellar evolution codes using parameterized approaches.
%: mixing length theory for convection, prescribed overshooting profiles at convective boundaries, and diffusive treatments of wave-induced transport. 
Although these models have been successful in reproducing many global stellar properties, they cannot capture the inherently three-dimensional, nonlinear, and anisotropic nature of fluid motions inside stars. These motions are governed by the fluid dynamics equations, which remain very expensive to solve numerically because of the range of time and length scales characteristic of stellar interiors. The MUltidimensional Stellar Implicit Code \citep[\music][]{Viallet2013, Viallet2016, Goffrey2017} is developed to face these challenges. It solves the dynamics equations of a fully compressible fluid coupled with thermal diffusion, employing an implicit time-integration scheme that allows the efficient evolution of low Mach flows over long timescales \citep{Viallet2011}. In this proceeding, we review the recent applications of the \music code.

%One of the main challenges when modelling stellar interiors is the range of length and time scales involved. The characteristic lengths range from the dissipation scale in the convection zone, which is a few centimetres \citep{Canuto2009}, to approximately one solar radius for solar-like stars, which is $R_{\odot} \simeq 6.957\times 10^8$ m, and even multiple times this value for more evolved or higher mass stars. Similarly, the characteristic time scales involved range from a few minutes for acoustic waves to several billion years for a typical stellar lifetime. 

\section{Equations solved by MUSIC}
%%-------------------------

%\subsection{Equations solved by \music}
%%---------------------------------
The equations for mass, momentum, and internal energy conservation in \music are written in the form:
\begin{equation}
\frac{\partial \rho}{\partial t} = - \vec{\nabla} \cdot (\rho \vec \velvec),
\label{eq:mass}
\end{equation}

\begin{equation}
\frac{\partial\rho \vec\velvec}{\partial t} = - \nabla \cdot (\rho \vec \velvec \otimes \vec \velvec) - \vec{\nabla}p + \rho\vec{g} - 2 \rho\vec{\Omega} \times \velvec - \rho\vec{\Omega} \times \left( \vec{\Omega} \times \vec{r} \right),
\label{eq:momentum}
\end{equation}

\begin{equation}
\frac{\partial \rho e}{\partial t} = - \vec{\nabla} \cdot (\rho e \vec\velvec) - p \vec{\nabla} \cdot \vec \velvec - \vec{\nabla} \cdot \vec{F_r} + \rho \epsilon_{\rm nuc},
\label{eq:energy}
\end{equation}
where $\rho$ is the density, $e$ the specific internal energy, $\vec \velvec$ the velocity field in the rotating frame, $p$ the gas pressure, $\Omega$ is the rotational angular frequency, $\epsilon_{\rm nuc}$ is the specific energy released by nuclear burning and $\vec{g}$ the gravitational acceleration. 
The hydrodynamical simulations run for this work assumes spherically symmetric gravitational acceleration, $\vec{g} = - g \vec {\rm e}_r$, which is computed within each implicitly solved timestep as:
\begin{equation}
    g(r) = 4\pi \frac{G}{r^2} \int^r_0 \overline{\rho}(u)u^2 {\rm d}u,
\end{equation}
with $\overline{\rho}(r)$ radial density profile calculated by averaging over the sphere of radius $r$.
%given by
%\begin{equation}
%    \overline{\rho}(r) = \AvgS{\rho}
%\end{equation}
%where the operator $\AvgS{.}$ is an angular average over %the whole unit sphere, defined as
%\begin{equation}
%    \left< h \right>_{\mathcal{S}} \coloneqq \frac{1}{4\pi} \int_\mathcal{S} h(\theta, \phi) \, 2\pi \sin \theta {\rm d} \theta {\rm d} \phi.
%    \label{eq:angular_av}
%\end{equation}

For simulations of stellar interiors, the major heat transport mechanism that contributes to thermal conductivity is radiative transfer characterised by the radiative flux  $\vec{F_r}$, given within the diffusion approximation by

\begin{equation}
\vec{F_r} = - \frac{16 \sigma T^3}{3\kappa \rho} \vec{\nabla} T = - \chi \vec{\nabla} T,
\label{eq:radiative_flux}
\end{equation}
with $\kappa$ denoting the Rosseland mean opacity of the gas, $\sigma$ the Stefan–Boltzmann constant and $\chi$ the thermal conductivity respectively.
\textsc{music} incorporates realistic stellar opacities \citep[OPAL, ][]{Iglesias1996} for metallicity in the range Z=0.0 to Z=0.04, and equations of state \citep[OPAL EOS, ][]{Rogers2002} suitable for describing stellar interiors.

%%---------------------

\section{Studying the interior of stars with MUSIC}
\music is designed to study the internal dynamics of stars, with a particular focus on convection, convective boundary mixing (CBM) and waves. These processes are multidimensional, non-linear and anisotropic. Their description based on 1D descriptions, as used in stellar evolution models, is therefore very uncertain.

\subsection{Convection}
\begin{figure}[ht!]
 \centering
 \includegraphics[width=0.8\textwidth,clip]{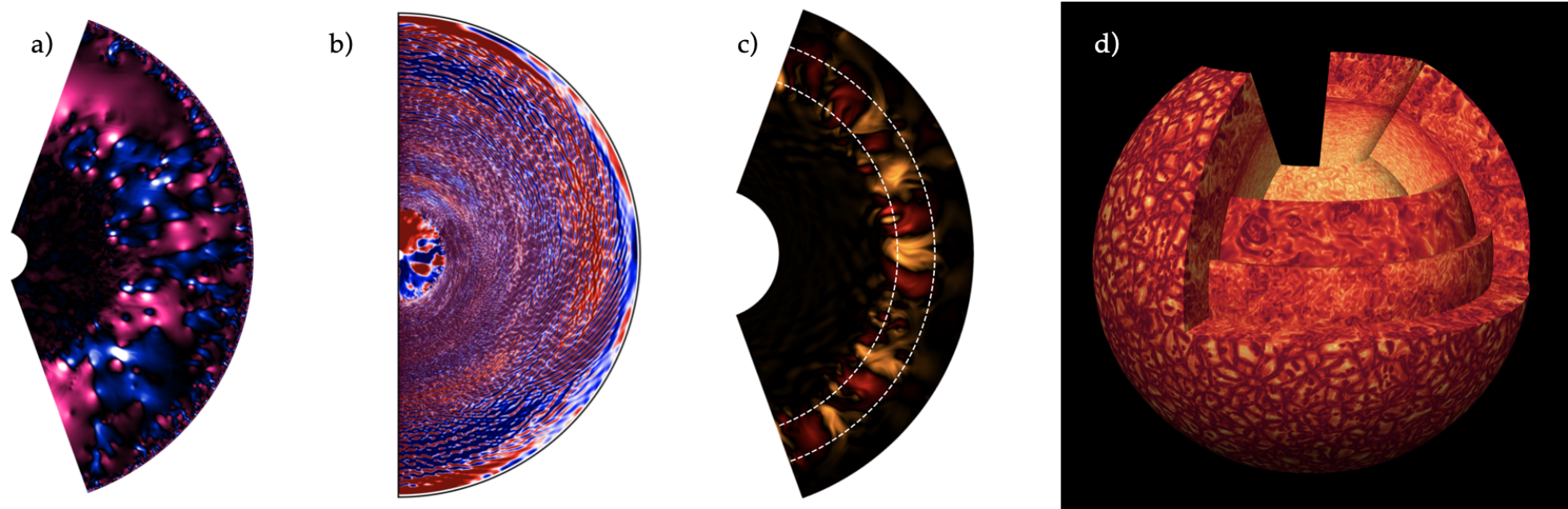} 
%% Note the ABSENCE of the extension .pdf  !
  \caption{Visualisation of the radial velocities in \music simulations for: a) 2D young Sun \citep{Dethero2024}, b) 2D 5 $M_{\odot}$ star \citep{LeSaux2023}, c) 2D Cepheid variable star \citep{Stuck2025} and d) 3D solar model (Vlaykov et al. in prep).}
  \label{fig:snapshots}
\end{figure}

Convection is a universal feature of stellar interiors. Depending on the structure of the star, it may occur at the centre in the form of a convective core, near the surface as a convective envelope, or even as a convective shell embedded within a large radiative layer. These three different configurations are illustrated in Fig. \ref{fig:snapshots}. 

In these regions, turbulent macroscopic fluid motions transport heat and chemicals very efficiently. In one-dimensional models, this is usually described using the mixing length theory \citep[MLT; e.g.][]{Bohm-Vitense1958}. However, this simplified description remains incomplete, and as such global stellar hydrodynamical simulations are needed to capture the relevant large-scale dynamics. The stratification in such zones impacts the properties of convection, particularly close to the surface of solar-like stars, where this stratification becomes very steep. In \music simulations of a 1$M_{\odot}$ model, \citet{Vlaykov2022} show that including more or less outer layers of the outer envelope significantly affects convective motions. Extending the outer boundary of the simulation domain by only a few per cent of the stellar radius tends to increase the velocity and temperature perturbations linked to convection. This study is based on 2D simulations, as their reasonable computational cost allows us to make parametric studies. 
In \citet{Pratt2020} the comparison of 2D and 3D \music simulations highlights that in the 3D case, convective velocities are lower than in 2D. Nevertheless, the large-scale dynamics in 2D and 3D simulations appears to be similar.
In a recent study, \citet{Dethero2024} extended this work, lookin in more detail at the shape of convection in stars of different masses and evolutionary stages. This shape, or geometry, of convective motions is crucial for 1D parameterizations of convective overshooting as these are usually described using a filling factor of convective inflows \citep[e.g.][]{Schmitt1984, Zahn1991, Brummell2002, Kapyla2017}. This filling factor defines how much of the fluid motions are moving inward in the star \citep{Stein_Nordlund1989}. \citet{Dethero2024} show that the flows in stars with convective envelopes are asymmetric close to the photosphere, but tend to become symmetric at the bottom of the convective zone. By looking at the width and number of plumes travelling inward (in convective envelope) or outward (in core convection), the authors are able to find different statistical properties between 2D and 3D convection. However, these do not fully explain why the overshooting observed in 2D and 3D simulations has the same extent.

\subsection{Convective Boundary Mixing}
%As of today, CBM remains one of the most uncertain aspects of stellar physics, yet it plays a crucial role in the evolution of stars. 
At the interface between convective and radiative zones, turbulent motions can overshoot beyond the formally unstable region and mix material into adjacent stable layers. Depending on the balance between turbulent convection and stabilizing stratification, CBM can be described as overshooting, when the temperature gradient in the penetration region is kept radiative, or penetrative convection, where the extended region becomes nearly adiabatic.
In 1D stellar models, MLT predicts that convective motions stop at the interface with adjacent radiative zones. To include the effect of CBM, simple prescriptions have been developed, based on a diffusion coefficient $D(r)$ to characterise the mixing in the penetration region. While these approaches offer some flexibility, $D(r)$ is difficult to estimate from theoretical arguments or calibrate from observations. 
Thus, there is no universal agreement on the shape of D(r). Simulations have suggested exponential \citep{Hurlburt1994, Jones2017}, step \citep{Lecoanet2015} or Gaussian \citep{Korre2019} profiles.
Using 2D \music simulations of a young Sun, \citet{Pratt2017} developed a statistical method based on extreme values theory and assumed that rare, extreme penetrating events -- and not the average -- characterise the relevant penetration depth, as they contribute to mixing on the long term. This led to a statistical description based on a Gumbel distribution. 
Based on this statistical method, \citet{Baraffe2021} demonstrated that penetrative motions induce shear and thermal mixing in the penetrative region of solar-like model, modifying the local thermal background. The associated local heating impacts the sound speed profile below the convective boundary, which tends to reduce the disagreement usually reported between solar models and helioseismic observations \citep{Basu2009}. Nevertheless, this kind of thermal effect is challenging to study in global stellar simulations. These simulations remain computationally expensive, so achieving thermal relaxation is unaffordable with current computing resources. To overcome this challenge, a common technique, known as boosting, artificially increases the luminosity and thermal diffusivity of a model by several orders of magnitude \citep[e.g.][]{Brun2011, Rogers2013, Andrassy2024}. While boosting accelerates relaxation, it also alters physical processes: enhanced heating in the penetrative region modifies overshooting and wave dynamics, meaning that boosted models cannot be treated as simple “accelerated” versions of a reference model \citep{Baraffe2021, LeSaux2022}.

Recently, \citet{Stuck2025} ran 2D \music simulations of a Cepheid model to study CBM in a radiative shell between two convective regions. They developed a mixture model to be able to tell differences in the mixing of different parts of this radiative shell, and whether there were layers in the radiative shell that could be mixed by both the convective layer below and the convective layer above.

In stars with convective cores, CBM is particularly important because it determines how much extra fuel a star can get to extend its life in the main-sequence. Standard stellar models generally predict lifetimes that are too short for intermediate- and high-mass stars \citep{Johnston2021}, and this is often corrected by artificially increasing the size of the penetration zone $l_{\rm ov}$. Observations suggest a mass dependence of $l_{\rm ov}$ \citep[e.g.][]{Claret2019, Castro2014}. Using 2D \music simulations of stars between 3 and 20 $M_{\odot}$ \citet{Baraffe2023} derived a scaling law, $l_{\rm ov} \propto L^{1/3}r_{\rm conv}^{1/2}$, with $L$ the luminosity of the star and $r_{\rm conv}$ the size of the convective core. Implementing this relation in 1D models improves agreement with observed main-sequence durations. Yet, for stars with $M > 10 M_{\odot}$, the predicted overshooting does not explain the width of the observed main-sequence in a Hertzsprung-Russel diagram, pointing to additional mixing processes above the core. By performing a similar work for overshooting beneath convective envelopes, \citet{Pratt2025} found some interesting similarities with \citet{Baraffe2023}, in particular a scaling law $l_{\rm ov} \propto L^{1/4}$.

%Compared to observations, stellar evolution models predicts a too short lifetime for intermediate and massive stars \citep{Johnston2021}. To improve agreement with observations, the overshooting distance $l_{\rm ov}$ is adjusted to increase the size of the convective core. Recent studies have revealed that $l_{\rm ov}$ depends on stellar masses \citep[e.g.][]{Claret2019,Castro2014}. In \citet{Baraffe2023}, we study this mass dependency with 2D hydrodynamical simulations of stellar models with masses between 3 and 20 $M_{\odot}$. We derive a scaling law $l_{\rm ov} \propto L^{1/3}r_{\rm conv}^{1/2}$, with $L$ the luminosity of the star and $r_{\rm conv}$ the size of the convective core. When implemented in 1D stellar evolution models, this scaling law tends to give better agreement with observations. However, for stars with mass $M > 10 M_{\odot}$, the predicted value of $l_{\rm ov}$ still underestimates the duration of the main-sequence, hinting that other mechanisms are probably causing mixing above the core.

\subsection{Waves}
Turbulent motions in convective zones excite waves that propagate through stellar interiors. In non-rotating, non-magnetic stars, the two main kinds of wave generated are acoustic, restored by pressure gradients, and internal gravity waves (IGW), restored by buoyancy. Theory has long suggested that IGW are excited by convection through Reynolds stresses \citep{Stein1967, Press1981, Goldreich1990, Lecoanet2013} and by penetrative convection \citep{Rieutord1995, Montalban2000, Pincon2016}. Yet, despite decades of work, their excitation spectra remain poorly constrained.
%Nevertheless, despite decades of research, convectively excited IGW in stellar interiors remain elusive, even in the solar case. One of the main reason for that is because we still do not know the actual excitation spectra of IGW, thus the need for multidimensional hydrodynamical simulations. 
Using \music, \citet{LeSaux2022} showed that IGW energy flux is consistent with theoretical expectations for waves excited simultaneously by Reynolds stress and convective penetration and that their damping follows the predictions of radiative damping by \citet{Press1981}. Extending this to stars with convective cores, \citet{LeSaux2023} highlighted that wave damping is very sensitive to the radiative diffusivity profile used in simulations, particularly in the upper layers of a star. When studying whether IGWs can reach all the way to the photosphere, it is therefore crucial to use realistic radiative diffusivity profiles. 
%and then be detected by photometric observations. 
Recent observations suggest that IGW do indeed reach the surface, which is proposed as an explanation for the Stochastic Low Frequency variability in O and B type stars \citep{Bowman2019, Bowman2020}, but these are contradicted by theoretical work \citep{Lecoanet2021}. Using 2D simulations with realistic thermal diffusivity profiles, \citet{LeSaux2023} suggested that low-frequency waves are radiatively damped before reaching the surface, a result corroborated by 3D simulations \citet{Anders2023}. 
%These results were recently questioned by (REF).
%In addition, physical phenomena such as age and rotation can impact the excitation and propagation of IGW. 
More recently, \citet{Morison2024} extended this analysis to stars at a more advanced stage on the main sequence and showed that the strong stratification left behind by core contraction reduces the efficiency of wave excitation and strongly damps them. This makes their detection at the surface more unlikely.

%in which the contraction of the core creates a strongly stratified region. We demonstrated that this stratified region reduces the efficiency of wave excitation and strongly dampens them, decreasing the probability of observing them at the surface of advanced stars on the main sequence.

On Earth, IGW are known to mix materials in the atmosphere and in the ocean through nonlinear effects, such as wave breaking \citep[e.g.][]{Lindzen1981}. This is less clear in stellar interiors. Chemical transport has been predicted by waves damped by radiative diffusion \citep{Press1981} or by wave-induced shear \citep{GLS1991}. In hydrodynamical simulations,  mixing in radiative regions has been linked to IGW \citep{Rogers2017, Varghese2023}, but it remains unclear what the actual physical mechanism responsible for wave mixing is. In 2D simulations of a 20 $M_{\odot}$ star model, \citet{Morton2025} showed that the derived mixing rate in radiative regions cannot be explained by the theoretical predictions of \citep{Press1981} and \citep{GLS1991}. This study also highlighted the challenge of quantifying wave mixing with Lagrangian tracer particles.

Waves also provide the only probe of stellar interiors thanks to global oscillation modes. This is the field of asteroseismology. Using acoustic modes, this method was exceptionally successful in measuring the internal structure and rotation profile of the Sun down to 20\% of the total radius \citep[see the reviews by][]{JCD2021, Howe2009}.
%In the Sun, this method was exceptionally successful at measuring the internal structure and rotation profile of our closest star \citep[see the reviews by][]{JCD2021, Howe2009}. These measures were obtained using acoustic modes and revealed the solar interior down to 20\% of the total radius. 
However, the innermost layers remain currently out of reach, as the acoustic modes do not propagate so deep. Using wave topology and \music simulations, \citet{LeSaux2025} confirmed the theoretical work of \citet{Leclerc2022}, which describes the fundamental $f$-mode at low angular degrees $\ell$ as a compressible bulk mode. More importantly, \citet{LeSaux2025} also predicted the existence of mixed $f$/$g$ modes in the Sun. These modes, with simultaneous amplitude in the core and in the envelope, offer a very promising pathway for probing the solar core. This work illustrates the potential of combining global stellar simulations with observational seismology to explore the interior of stars. It is currently extended to mixed p/g modes in low-mass subgiants and red giants to study their amplitudes and understand their efficiency to transport angular momentum (de Vries et al. in prep).

%which have the particularity of having amplitude simultaneously in the core and in the envelope. This result confirms the theoretical work of \citet{Leclerc2022}, which describes the f mode at low angular degree $\ell$ as compressible bulk mode. More importantly, the coupling between f and g modes might provide the best changes to probe the solar core in the future. This study by \citet{LeSaux2025} initiated a path towards improving the synergy between observational seismology and global stellar simulations. 

\section{Future developments}
\begin{figure}[ht!]
 \centering
 \includegraphics[width=0.8\textwidth,clip]{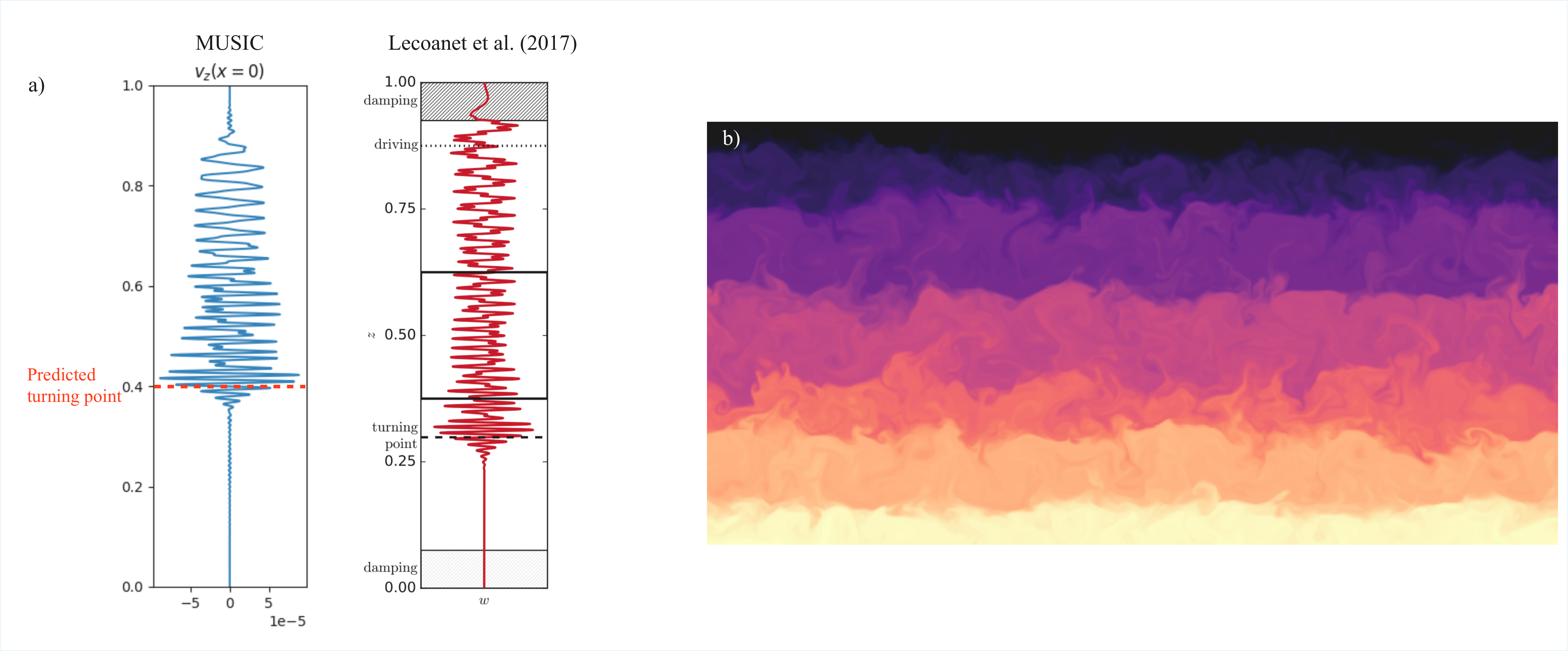} 
%% Note the ABSENCE of the extension .pdf  !
  \caption{New applications of \music. Left: conversion of an IGW into an Alfvénic wave in the presence of vertical magnetic field. Right: Double-diffusive convection with unstable composition gradient and stable temperature gradient.}
  \label{fig:future}
\end{figure}

Taken together, these studies illustrate how \music simulations are progressively bridging the gap between simplified 1D prescriptions and the high complexity of stellar interiors. In addition, fluid dynamics in stars is influenced by rotation, magnetic field, and microphysics such as opacity and equation of state.
To include these effects, \music is in constant development. Current studies now include rotation, in an effort to understand the impact of rotation on the amplitude of acoustic modes in solar-like stars (Le Saux et al., in prep). Ideal and non-ideal MHD is also implemented in \music. Using box-in-a-star simulations in Cartesian geometry, we are currently quantifying the impact of magnetic fields on IGW. As predicted by \citet{Fuller2015} and \citet{Lecoanet2017}, we observe in \music the conversion of IGW into Alfvénic waves for critical values of magnetic field. This is illustrated in Fig. \ref{fig:future}.
We are also aiming at expanding the \music range of physical applications, particularly towards giant planet interiors. Indeed, despite a very different regime in terms of Prandtl number, internal fluid dynamics of stars and giant planets share similar properties. However, the internal structure of giant planets remains a mystery. The space missions Cassini and Juno have revealed that the interiors of Saturn and Jupiter are not fully convective \citep{Debras2019, Mankovich2021}. In order to reproduce the observational constraints on gravitational moments, double-diffusive convection layers must exist in these giant planets.
We are currently investigating the formation and sustainability of such layers based on Cartesian box simulations that include a stable composition gradient and an unstable temperature gradient (see Fig. \ref{fig:future}). The next steps will include stratification, spherical geometry, and a realistic equation of state for planetary interiors.

%%
%% Example of single figure
%% PLEASE ONLY PROVIDE PDF FIGURES
%%

%%
%% Example of two figures side by side
%%
%\begin{figure}[ht!]
% \centering
 %\includegraphics[width=0.48\textwidth,clip]{Surname_SXX_fig1}%      
 %\includegraphics[width=0.48\textwidth,clip]{Surname_SXX_fig2}      
%% Note the ABSENCE of the extension .pdf  !
%  \caption{{\bf Left:} Caption of the left panel. {\bf Right:} Caption of the right panel.}
%  \label{author1:fig2}
%\end{figure}

%\section{Conclusions}
%%--------------------

% Optional acknowledgements
% -------------------------
\begin{acknowledgements}
\textit{Acknowledgements.} ALS acknowledge support from the European Research Council (ERC) under the Horizon Europe programme (Synergy Grant agreement 101071505: 4D-STAR). While partially funded by the European Union, views and opinions expressed are however those of the author only and do not necessarily reflect those of the European Union or the European Research Council. Neither the European Union nor the granting authority can be held responsible for them. Part of this work was performed under the auspices of the U.S. Department of Energy by Lawrence Livermore National Laboratory under Contract DE-AC52-07NA27344.  LLNL-PROC-2012036
\end{acknowledgements}

\bibliographystyle{aa}  % A&A bibliography style file (aa.bst)
\bibliography{Le_Saux_S07} % your references in file: Yourfile.bib

\end{document}